\documentclass[preprint]{aastex}

\shorttitle{IRS16SW Variability}
\shortauthors{DePoy et al}

\begin{document}

\title{The Nature of the Variable Galactic Center Source IRS16SW}

\author{D. L. DePoy, J. Pepper, Richard W. Pogge, Amelia Stutz, M. Pinsonneault, \& K. Sellgren}
\affil{Department of Astronomy, The Ohio State University}
\affil{140 West 18$^{th}$ Avenue, Columbus, Ohio 43210-1173\\
depoy@astronomy.ohio-state.edu, pepper@astronomy.ohio-state.edu,
pogge@astronomy.ohio-state.edu}


\begin{abstract}
We report measurements of the light curve of the variable Galactic Center
source IRS16SW. The light curve is not consistent with an eclipsing binary
or any other obvious variable star. The source may be an example of a 
high mass variable predicted theoretically but not observed previously.
\end{abstract}




\section{Introduction} \label{sec:intro}

The star cluster at the Galactic center (GC) is unique in the Milky Way.
It is composed of a mixed population of young and old stars in a very
dense cluster around the central supermassive black hole.
Various efforts to study the members of this cluster have involved
imaging and spectroscopic observations (e.g. \citet{bsdp96} 
and \citet{krabbe95}); all in the infrared since
the extinction to the GC is A$_V\approx$ 30 mag \citep{bn68}. \citet{tcb98}
and \citet{bsdp96} also discussed the variability of stars as probes of
the stellar population content, although the data sets involved had
sparse temporal sampling. More recently, \citet{ott99} used the results
from seven years of observing (roughly 6-7 nights per year) to identify
many variable stars in the GC, including the discovery that one of the sources
very near the GC, IRS16SW, is a short period variable.

In 1999 we began a long-term project to monitor the GC. We obtained
infrared images of the GC for many nights in subsequent observing seasons. The
goals are to identify all the variable stars in the GC and use the
statistics to better understand the star formation history of the
region. The data collection for this project is on-going and we expect
to produce the full results eventually. Here we report on the light curve of
IRS16SW from images taken during the 2001 observing campaign.

We find that IRS16SW may be a representative of a new class of regularly pulsating
massive stars. Pulsating stars are critical for the determination
of the extragalactic distance scale, Cepheid variables in 
particular (see Gibson et al. 2000). However, even Cepheids are difficult
to detect at large distances. Massive stars can have bolometric luminosities
hundreds of times larger than Cepheids and can
therefore be seen over much larger distances. The behaviour of massive
stars is intrinsically complex, however, so their use as standard candles
has been limited (see Kudritzki, Bresolin, \& Przybilla (2003) 
and Kudritzki et al. (1999) for additional discussion). Thus, 
the combination of regular pulsational brightness variation and
high luminosity that we observe in IRS16SW could have significance
beyond simple interest in this particular source.


\section{Observations, Data Reduction, and Results}\label{sec:obs}

Images of the Galactic center in the H (1.65 $\mu$m) and K (2.2 $\mu$m)
bands were obtained at the CTIO/Yale 1m telescope using the facility
optical/infrared imager \citep[ANDICAM; see][for a description of the
instrument]{depoy02}.  On the CTIO/Yale 1m telescope
the ANDICAM's infrared camera has a pixel scale
of 0\farcs22 pix$^{-1}$, which provides a total field of view of
225\arcsec\ on the 1024$\times$1024 HgCdTe array.

The first images for our 2001 observing campaign were obtained on UTC
2001 May 20 (HJD 2452049.5); additional images were obtained on every
usable night through UTC 2001 November 2 (HJD 2452216.5).  On each
night, we acquired seven individual images of the Galactic Center, each
slightly offset from the others, in each of the H and K filters. 
The individual H images were 30 second exposures; the individual K
images 10 second exposures. The groups of seven images required roughly
4 minutes at H and 2 minutes at K to obtain.
Between the H and K images we acquired similar sequences of images of a
sky position several degrees from the GC.  The sets of images were flat
fielded using dome flats and then shifted and combined to create single
H and K images for that night; a similarly processed sky image was then
used to provide the sky subtraction for each filter.

The final nightly images were trimmed to a size of 512$\times$512 pixels
providing a field of view of 112 arcseconds centered approximately on
the Galactic Center.  Upon inspection, some of the images were found to
be of poor quality; typically due to poor seeing, bad telescope focus or
tracking, or excessive wind shake. 89 K images and 83 H
images were retained for further analysis.  Of these, 75 images in each
band are on the same night, providing contemporaneous measurements of
the $H-K$ colors.

The images of the Galactic center at H and K are extremely crowded
(roughly 5 bright sources within $\sim$2 arcseconds in the IRS16
complex; see \citet{ghez98} and \citet{eckhart02}).  Further, the median
seeing in our images is $\sim$1\farcs1, but ranges from a minimum of
0\farcs94 to a maximum 2\farcs4.  Therefore, simply performing aperture
photometry on IRS16SW does not produce an adequately accurate light
curve.  Instead, we used the {\sc ISIS} image subtraction package
\citep[see][]{al98,alard00} to analyze the images.

We used the ISIS package to combine four of the best-seeing images in
each band into a template image, and then subtracted the template from
each individual image, after convolving the template to the
seeing in each image.  The resulting subtracted images allowed us to
measure all variable sources in the observed field, including IRS16SW.

We photometrically calibrated the light curves by comparing the
brightness of IRS16SW in the template image to the brightness of
isolated stars in the frame. These stars varied by $\la 0.5\%$ over the
course of the 2001 observing season; the brightness of these stars were
assumed to be the same as reported by \citet{bsdp96}. Intercomparison of
10 of these stars suggests that the absolute calibration of the
photometry is accurate to $\sim10\%$.  We note that \citet{ott99} find
the mean K brightness ($m_{K}$) of IRS16SW to be $\sim$9.61 mag;
we find $m_{K}$ $=$ 9.5$\pm$0.1 mag.  The relatively insignificant
difference could be due to
slight differences in the effective wavelength of the filters used in
the two data sets, since IRS16SW is very red. Our measurement is also
consistent with previous high angular resolution measurements
of $m_{K}$ (e.g. Simon et al. 1990,
Simons, Hodapp, \& Becklin 1990, DePoy \& Sharp 1991, and 
Blum, Sellgren, \& DePoy 1996).

We find that the mean H$-$K color of IRS16SW is 2.6$\pm$0.15 mag. 
Previous determinations of the H$-$K color of IRS16SW include
2.00$\pm$0.07 mag \citep{bsdp96},
2.05 mag (Rieke, Rieke, \& Paul 1989),
2.1$\pm$0.3 mag (Simons, Hodapp, and Becklin 1990),
2.4 mag (Krabbe et al. 1995), and 2.8$\pm$0.2 (DePoy \& Sharp 1991). 
Our measurement is generaly consistent with these previous 
determinations, although it is somewhat redder than the most
precise (e.g. Blum, Sellgren, \& DePoy 1996). The marginally
significant ($\sim$4$\sigma$) difference between our determination 
of the H$-$K color of IRS16SW and that of \citet{bsdp96} suggests
that our H-band magnitudes may contain systematic uncertainties.
Note that \citet{bsdp96} measured the H$-$K color on a 
single night (13 July 1993), so the variability of the color
of IRS16SW (see below) does not account for the difference.

\begin{figure}
\plotone{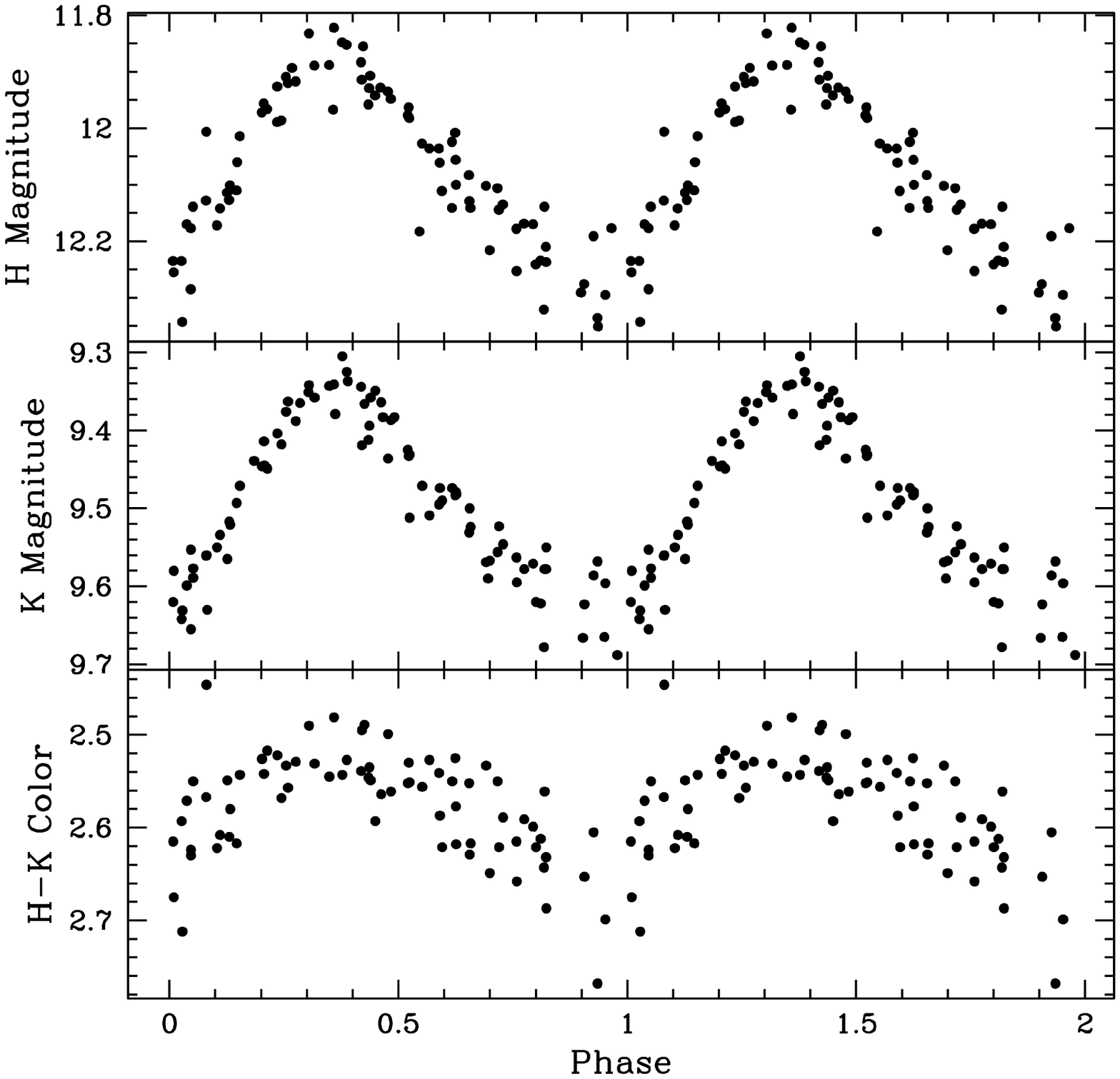}
\caption{
Light curve of IRS16SW in 2001. Data from $\sim$85 nightly images
taken over a period of 167 days are shown versus phase in a 9.725 day
cycle. Note that the color of the source is bluest at 
maximum light (H$-$K$\approx$2.52 mag) and reddest at minimum 
light (H$-$K$\approx$2.68 mag).
\label{fig:f1}
}
\end{figure}

We searched for the periodic signal in the H and K light curves of
IRS16SW using the Multiharmonic Analysis of Variances (ANOVA)
period-search algorithm described by \citet{sc96}.  We used a C-language
implementation based on a program provided by Christophe Alard for this
analysis.  Two harmonics were required to get a good periodogram fit,
giving a period of variability of 9.725$\pm$0.005 days, consistent with
that reported by \citet{ott99}.  The period found for each of the H and
K lightcurves was consistent to within the stated uncertainty and
we see no evidence for higher frequency overtones.
Figure~\ref{fig:f1} shows the phase-folded light curves at H and K for IRS16SW.

We note that the shape of the light curve is similar to that found
by \citet{ott99}. There are some differences, however. In particular, 
our light curve shows more continuous change in brightness than that
of \citet{ott99}; we see no evidence of any part of the light curve
that remains constant for a substantial fraction of the phase.

Also shown in Figure~\ref{fig:f1} is the color change in IRS16SW over the period of
variation.  The $H-K$ color of the source changes by a total of
0.16$\pm$0.03 mag (exclusive of systematic calibration uncertainties).  
The source appears bluest ($H-K\approx$2.52 mag) at
maximum light and reddest ($H-K\approx$2.68 mag) at minimum light.

\section{Discussion}\label{sec:discuss}

The shape and period of the IRS16SW light curve suggests it is either an
eclipsing binary star system or some variety of periodic variable star.
Careful examination of the properties of the light curve, however, rules 
out most of the familiar classes of explanations. Below we review and discuss
these possibilities and present evidence that IRS16SW may represent a
new class of regularly pulsating massive stars.

\subsection{Well-Known Pulsational Variables}\label{sec:pulvar}

\citet{ott99} presented the possibility that IRS16SW is a
Cepheid variable. They concluded that IRS16SW was not likely a Cepheid
on the basis of the light curve shape and apparent brightness of the
source at K. They also note that the spectrum of IRS16SW from
\citet{krabbe95} is not consistent with that expected from a Cepheid.
Our data shows that IRS16SW changes color by $\sim$0.18 mag ($H-K$) over the
course of its light curve. This is also inconsistent with the behaviour
of Cepheids, which typically show $<$0.05 mag H-K color change over their
periods \citep[see][]{welch84}. It therefore seems unlikely that IRS16SW
is a Cepheid variable.

Nonetheless, the shape of the IRS16SW light curve is reminiscent of pulsating
variable stars (i.e.  a sharper rise to maximum light followed by a
slower decline). One possibility is that IRS16SW is a $\beta$ Cephei
variable. These are high mass stars (6-30 M$_\odot$) that pulsate due to
the $\kappa$-mechanism caused by ion absorption peaked at {\it
T}$\approx$2$\times$10$^5$ K that excites fundamental mode oscillations
\citep[see][]{dx01}.  Known $\beta$ Cephei variables are early
B-type stars (except for HD 34656; \citet{pk98,fullerton91}). This may
be due to the very short length of time higher mass stars spend in the
appropriate part of the Hertzsprung-Russell Diagram or because the
pulsations in higher mass stars are of very small amplitude and
difficult to detect \citep{dp93,dx01}. Furthermore, observations of B-type
$\beta$ Cephei variables suggest that the photometric amplitude of the
variation decreases with wavelength \citep{heynderickx94}. Also, the
periods of $\beta$ Cephei variables are typically $<$1 day. Both these
are not particularly consistent with observations of IRS16SW, although
there are no observations of this class of variables in the infrared or
any modeling predicting their behaviour.

We are not aware of any other well-observed class of pulsating stars
that have characteristics similar to IRS16SW.

\subsection{Eclipsing Binary Systems}\label{binary}

IRS16SW does not appear to be an eclipsing binary. 
If IRS16SW is an eclipsing binary system, then the
lack of any secondary eclipse requires that either one of the systems is
invisible (and the period is $\sim$9.725 days) or that both components
have the same effective temperature (and the period is actually
2$\times$9.725$=$19.45 days). However, we detect a color change over the
period, which indicates that if there are two stars in the system then
one is cooler than the other or that both are detected. Logically, then,
IRS16SW cannot be an eclipsing binary system.

Ott et al. proposed that IRS16SW is an eclipsing binary and described two possible scenarios.  
The first was a contact binary system with a high-mass 
primary dominating the total light from 
the system eclipsed by a lower-mass companion in a 9.725 day orbit around the primary.  
In such a system, we would expect to see a relatively deep primary eclipse 
followed by a shallow secondary 
eclipse half an orbital phase later. This is not seen in 
our light curve (see Figure~\ref{fig:f1}), 
and so can be ruled out.  The second scenario was of a contact binary 
composed of two equal mass 
stars each contributing equally to the total light. Dividing the 
light equally between the stars 
gives them radii of 64\,R$_\sun$, implying minimum masses of 75\,M$_\sun$ 
each for two stars in contact 
given an orbital period of 19.450\,
days (2$\times$9.725\,days).\footnote{Ott et al. reported minimum 
masses of $>$150\,M$_\sun$ for this case, but it is clear that 
they inadvertently used an orbital period 
of 9.725\,days instead of 19.45\,days as required by an 
equal-mass/equal-radius eclipsing binary 
system in which the primary and secondary eclipses are identical}

\begin{figure}
\plotone{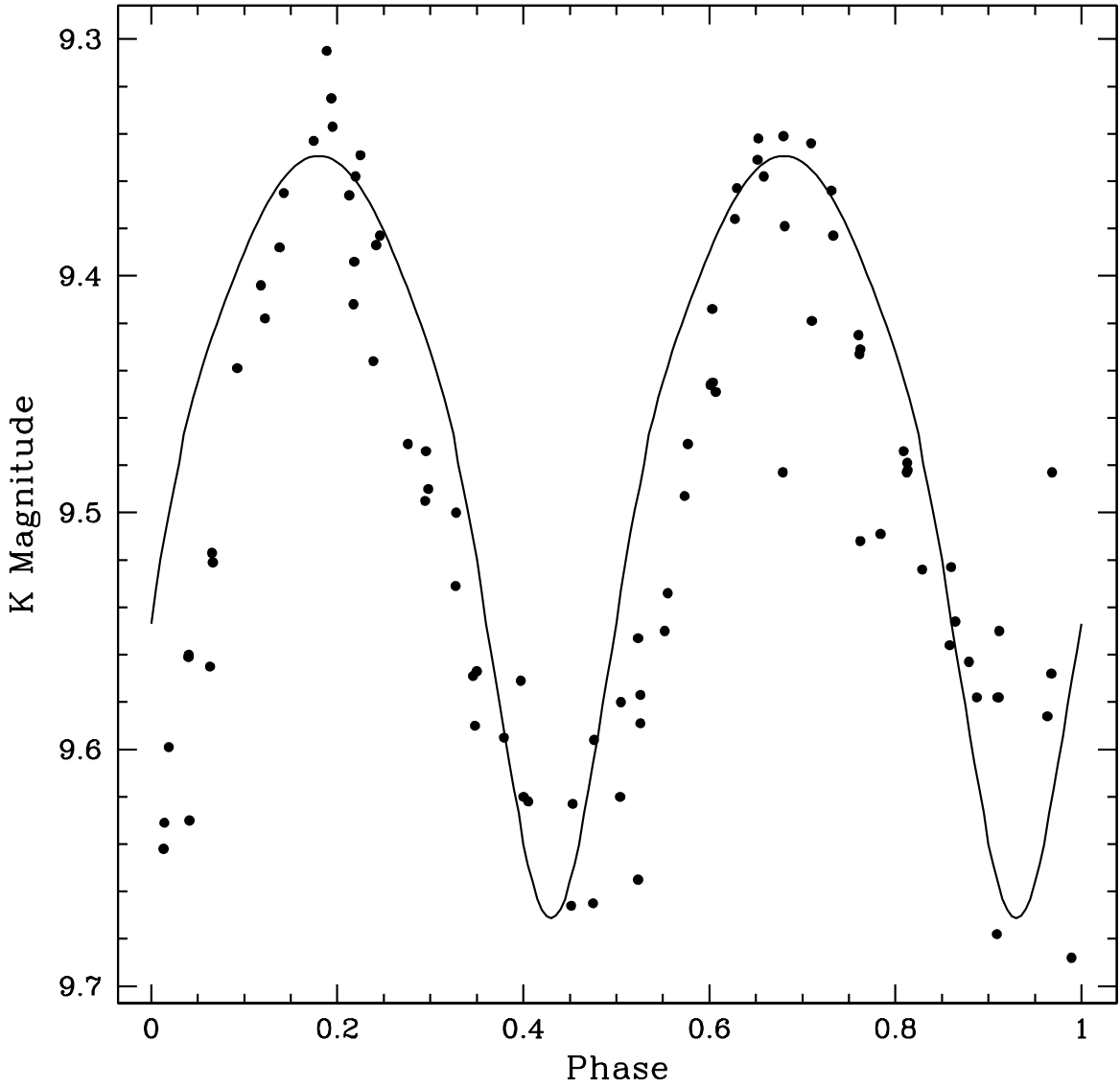}
\caption{
K-band light curve versus phase (P$=$19.45 days) of IRS16SW with 
eclipsing binary model discussed in the text
superimposed. The model does not represent the data well.
\label{fig:f2}
}
\end{figure}

We modeled the expected light curve for this equal-mass contact eclipsing binary system 
using a Wilson \& Devinney model \citep{wd71,w79,w90} (we used a 1998 version of the 
code generously provided by R.E. Wilson).  The Wilson \& Devinney model correctly 
accounts for all of the relevant physics in two stars sufficiently close together to be 
distorted by their mutual gravitational fields.  For this calculation we used two 
equal-mass stars with $T_1=T_2=24400K$ and $R_1=R_2=64$\,R$_\sun$ in a contact binary 
system with a circular orbit of semi-major axis $a=64$\,R$_\sun$ and a period 
of $P=19.450$\,days. The temperature and radius were selected based on models of
IRS16SW by Najarro et al. (1997) modified to account for the binarity of the
source (as suggested by Ott et al.).
The code modeled the stars as in contact and distorted 
by tides, using limb-darkening parameters typical of electron-scattering atmospheres 
expected in such high-mass, hot stars.  An orbital inclination relative to the line 
of sight of $i=65\deg$ produces a double eclipse with a depth of 0.3 mag, matching 
the amplitude of variability seen in our K-band light curve shown 
in Figure~\ref{fig:f1} (note that if $i=90\deg$, the eclipse depth 
should be exactly 50\% of the light, or 0.75 mag).  We note that we did not 
attempt to ``fit'' our light curve (other than by attempting to match the
observed amplitude of the light curve), rather we generated the predicted light 
curve for the equal-mass contact binary system described.  The model light 
curve is shown plotted over our observed light curve in Figure~\ref{fig:f2}.  
This model is clearly a poor match to the observed light curve. Further,
if the two stars have the same mass and evolutionary state, then there 
should be no systemic color change over the period; as mentioned above,
this is contrary to
the observed color change. Therefore, we rule out the equal-mass contact
binary scenario.

\subsection{A Massive Regularly Pulsating Variable?}\label{subsec:new}

If we rule out that IRS16SW is a known type of pulsating star or an eclipsing
binary system, then it must represent a new class of variable object. This
intriguing possibility is supported by recent work on internal models of
massive stars. In particular,
\citet{dg2000} found that linearly overstable pulsational modes
can develop into regular variability in radiation hydrodynamic simulations
of massive stars. In particular, for very high mass stars their models suggest
that these modes can cause very regular cyclic brightness variability with 
light curves similar to those of classic pulsating variables. The models suggest
that this variation is stable over reasonably long timescales.

The highest mass model presented by \citet{dg2000} is for a 
60 M$_\odot$, L $=$ 900,000 L$_\odot$, $T_{eff} = 18,000 K$ star (their model M60C).
The model predicts a regular pulsation with $P = 4.086$ days and peak-to-peak
bolometric brightness variation of 0.66 mag. The light curve looks qualitatively
like that we observe for IRS16SW: a relatively steep rise, followed by a somewhat 
slower fall. The model predicts no significant phase shift between different
filter passbands of features in the light curves, which \citet{dg2000} attribute 
to low heat capacity in the most superficial stellar layers. Color changes are
present, however, since the effective temperature of the star changes during the
pulsations. These aspects of the model are also consistent with our observations of
IRS16SW. Note that \citet{dg2000} simulated this model's pulsation for a timescale
of more than 20 years without seeing a change in the pulsational properties.

However, \citet{dg2000} modeled variations only in specific
optical passbands (UBVI) and their highest mass model has a
period less than half that of IRS16SW. Their models show that
the amplitude of variability decreases with wavelength throughout
the optical, with the largest amplitudes in the ultraviolet. The
amplitude of the M60C model in the I band is $\sim$0.2 mag, suggesting 
the variations in the near-infrared would be much smaller than
we observe in IRS16SW.
Furthermore, all their models with $M > 30 M_\odot$ show
secondary maxima caused by shocks waves in the atmospheres of the 
stars (a difference between the dynamical timescale of the
atmosphere and the pulsational period causes collapsing layers
to collide with already rerising deeper layers); we see
no evidence for these secondary maxima in our IRS16SW light curve.

\citet{dg2000} point out that there is no observational evidence
for pulsating very massive main-sequence stars with short periods. They
suggest this may be due to either a missing piece of physics in their
models, which would damp out the pulsations, or lack of appropriate
observational data, which would easily confuse the pulsations
with flickering or simply observational error. Their models 
suggest that both period and amplitude of variation
increase with mass, but this has not been modeled.

Thus, although it is attractive to speculate that IRS16SW is an example
of a massive and regularly pulsating star, additional theoretical work
is necessary before reliable conclusions can be made. To guide future 
theoretical efforts, a summary of the stellar parameters of IRS16SW 
is appropriate. The basic observational characteristic of the source
is that it is bright and red. As discussed in section 2, IRS16SW has mean
$m_{K}$ $=$ 9.5$\pm$0.1 mag; the reddening is A$_K$ $\approx$ 3 mag. This
corresponds to $M_{K}$ $\approx$ $-$8.2 mag (for an assumed 
distance of 8.5 kpc). The color of IRS16SW is dominated
even in the near infrared by the high extinction to the GC; the color does not
provide meaningful constraints on the characteristics of the star (that is,
the uncertainty in the colors and extinction do not place interesting
constraints on the spectral type of the object). IRS16SW also has strong
HI and HeI line emission; some of the HeI emission might be
due to CIII and NIII (see Krabbe et al. 1995). Najarro et al. (1997)
used the spectrum of IRS16SW to model the physical characteristics of
the star. Their model suggests that the star has a radius of $\sim$90 R$_\sun$,
a luminosity of $\sim$26 L$_\sun$, and an effective temperature of
$\sim$24,000 K. Their model also suggests IRS16SW has an outflow
of about 1.5$\times$10$^{-5}$ M$_\sun$ yr$^{-1}$. They conclude the star is
an evolved blue supergiant close to the evolutionary phase of 
Wolf-Rayet stars, although this conclusion may be tempered by future
theoretical work on the nature of the star's pulsations.


\section{Conclusions}\label{sec:conclude}

Observations of the galactic center region over roughly the entire 2001 observing 
season show that IRS16SW is a periodic variable star, confirming the
results of Ott et al. (1999). The period of the variation is 9.725$\pm$0.005 days
in both the H and K bands. 
This is the same period reported by Ott et al., demonstrating that IRS16SW has
had a period stable for at least the past $\sim$10 years.
There is a change in the $H-K$ color of IRS16SW over
this period of $\sim$0.16 mag.

The light curve shape and color change over phase demontrate that IRS16SW is
not an eclipsing variable star. Instead, the light curve is most similar
to that of a periodic pulsating star. However, the amplitude and color of 
the light curve and the lumiosity of IRS16SW are inconsistent with any known
type of pulsating variable. 

The observations are very roughly consistent with the predictions
for an unobserved class of periodically varying, high mass stars
made by \citet{dg2000}. Although the measured light curve resembles
the predictions, there were also serious differences (period, amplitude, etc.),
so the intriguing possibility that IRS16SW is the first of a new class
of high mass variable stars cannot be confirmed. 


\acknowledgements

We wish to thank Juan Espinoza and David Gonzalez for their dedication and
hard work at the telescope and the staff of the Cerro Tololo Inter-American
Observatory for their excellent support. ANDICAM was built with funds from
NSF grant AST-9530619 by the staff of the Imaging Sciences Laboratory of
the Ohio State University. We thank C. Alard for 
providing {\sc ISIS} and R. E. Wilson for supplying the program to
simulate contact binary light curves. Computing support provided by the
Ohio State University Department of Astronomy. 


%
%

\newpage

\end{document}